\documentclass[twocolumn,showpacs,preprintnumbers,amsmath,amssymb,prb]{revtex4}
\usepackage{amsfonts}
\usepackage{graphicx}

\begin{document}

\title{Lineshape of harmonic generation by metal nanoparticles\\ and metallic photonic crystal slabs}

\author{M.\,W. Klein}\email{Matthias.Klein@physik.uni-karlsruhe.de}
\author{T. Tritschler}
\author{M. Wegener}
\affiliation{Institut f\"ur Angewandte Physik, Universit\"at Karlsruhe (TH), D-76131 Karlsruhe, Germany}
\author{S. Linden}
\affiliation{Institut f\"ur Nanotechnologie, Forschungszentrum Karlsruhe in der Helmholtz-Gemeinschaft, D-76021
Karlsruhe, Germany}

%\date{\today}

\begin{abstract}
We study the linear and nonlinear-optical lineshapes of metal nanoparticles (theory) and metallic photonic crystal
slabs (experiment and theory). For metal nanoparticle ensembles, we show analytically and numerically that femtosecond
second- or third-harmonic-generation (THG) experiments together with linear extinction measurements generally do not
allow to determine the homogeneous linewidth. This is in contrast to claims of previous work in which we identify a
technical mistake. For metallic photonic crystal slabs, we introduce a simple classical model of two coupled Lorentz
oscillators, corresponding to the plasmon and waveguide modes. This model describes very well the key experimental
features of linear optics, particularly the Fano-like lineshapes. The derived nonlinear-optical THG spectra are shown
to depend on the underlying source of the optical nonlinearity. We present corresponding THG experiments with metallic
photonic crystal slabs. In contrast to previous work, we spectrally resolve the interferometric THG signal, and we
additionally obtain a higher temporal resolution by using 5\,fs laser pulses. In the THG spectra, the distinct spectral
components exhibit strongly different behaviors versus time delay. The measured spectra agree well with the model calculations.\\
(Submitted to {\it Phys. Rev. B} on February 28, 2005; published in {\it Phys. Rev. B} {\bf 72}, 115113 on
September 15, 2005; resubmitted to arXiv on \today)\\[5pt]
\copyright 2005 The American Physical Society
\end{abstract}


\maketitle

\section{Introduction}\label{SecIntro}

The linear-optical properties of metallic photonic crystal slabs (MPCSs) have recently attracted considerable attention
because they can be viewed as a simple model system of two coupled oscillators: (i) a Lorentz oscillator electronic
resonance couples to (ii) an electromagnetic resonance. (i) The electronic resonance comes about from charges which
accumulate at the surface of the metal nanostructures when exposed to the electric field of the incident light. These
charges induce a depolarization field that can either counteract or enhance the external electric field, depending on
the permittivity of the metal, hence depending on the frequency of light. The resulting resonance at the transition
point is the well-known {\it particle plasmon} or Mie resonance.\cite{KreibigBook} (ii) The electromagnetic resonance
is the {\it Bragg resonance} of the periodic arrangement with lattice constant $a$. Importantly, an appreciable
coupling between these two oscillators requires an additional slab waveguide, e.g., underneath the metal nanoparticles.
Therefore, the physics of metallic photonic crystal slabs is distinct from that of usual metallic gratings, which have
been discussed extensively many years ago.\cite{NeviereBook} Tailoring the waveguide parameters allows one to control
the coupling strength, since the coupling arises from the spatial overlap of the plasmon- and waveguide-mode fields.

Two-dimensional MPCSs were first discussed in Ref.\,\onlinecite{LindenPRL01} employing gold nanoparticles on a
dielectric waveguide. Later,\cite{ChristPRL03} gold nanowires showed even more pronounced effects. In the latter
structures, the coupling of the incident light to the particle plasmon resonance can conveniently be switched on and
off via the polarization. If the electric field vector is oriented perpendicular to the nanowires (TM polarization), a
pronounced depolarization field arises, giving rise to a strong optical resonance. In contrast, if the electric field
vector is along the wire axis (TE polarization), the depolarization factor is zero and one rather gets a Drude-type
response of the metal.

More recently, nonlinear-optical experiments on MPCSs have been presented.\cite{ZentgrafPRL04} These were interpreted
along the lines of similar experiments\cite{LamprechtAPB99} performed on metal nanoparticle ensembles on a substrate
surface without a slab waveguide. Thus, our initial motivation was to continue with experiments along these lines and
obtain additional information from nonlinear-optical experiments.

This paper contains both theory and experiments and is organized as follows. In Sec.\,\ref{SecEnsembles} we start by
discussing the nonlinear optics of metal nanoparticle ensembles with inhomogeneously broadened plasmon resonances. We
derive rigorous analytic results for excitation with $\delta$ pulses and Lorentzian inhomogeneous broadening.
Furthermore, we present numerical results for a Gaussian inhomogeneous broadening and finite-duration optical pulses.
For second- (SHG) and third-harmonic generation (THG), we find that the interferometric nonlinear response exclusively
depends on the total linear-optical linewidth of the particle plasmon, i.e., one cannot obtain any information on the
relative contributions of homogeneous and inhomogeneous broadening, respectively. This finding is in {\it striking
disagreement} with the claim of Ref.\,\onlinecite{LamprechtAPB99} that interferometric SHG together with linear-optical
measurements can differentiate between homogeneous and inhomogeneous broadening. Reference\,\onlinecite{LamprechtAPB99}
has been the basis of much if not most of the work that followed in this field.\cite{LamprechtPRL00, LamprechtPRL99,
AusseneggPRB03, TraegerPRL00, TraegerAPB99, ZentgrafPRL04, WeidaJOSAB00} The discrepancy is traced back to a simple
technical mistake in that work,\cite{LamprechtAPB99} where the authors have not properly differentiated between the
contributions of second-harmonic generation on the one hand and optical rectification on the other hand. Indeed,
optical rectification (OR), self-phase modulation (SPM), or four-wave mixing (FWM) would allow for such a
differentiation. Next, in Sec.\,\ref{SecLinOpt}, we discuss the linear-optical properties of two coupled oscillators
(representing the particle plasmon and waveguide resonance). In contrast to frequent belief, the optical response of
two coupled classical {\it damped} Lorentzian oscillators does {\it not} correspond to that of two new effective
Lorentzian oscillators. Generally, one rather gets Fano-like lineshapes in the linear-optical spectra. In
Sec.\,\ref{SecNonlinOpt} we discuss the nonlinear-optical signals from two coupled oscillators. We show that signatures
of interferometric THG depend on the source of nonlinearity. Our theoretical analysis thus yields additional insights
compared to the discussion in Ref.\,\onlinecite{ZentgrafPRL04}. The parameters of the presented numerical calculations
are chosen to allow for a direct comparison with our experimental results, which are presented in
Sec.\,\ref{SecExpmts}. Compared with previous work, our experiments are distinct in two aspects, (a) and (b). (a)
First, we use 5\,fs optical pulses, which are within the range of the anticipated particle plasmon decay times of
$0.7-9\,\rm fs$.\cite{FeldmannPRL02} Previous work used pulses of 13\,fs duration and
longer.\cite{LamprechtAPB99,ZentgrafPRL04} (b) As usual in ``time-resolved spectroscopy,'' indirect information on the
temporal behavior is obtained by exciting the sample with a pair of time-delayed pulses, e.g., in pump-probe or
transient four-wave mixing experiments. It is known that additional information can often be obtained by spectrally
resolving the probe beam or the diffracted beam. In analogy, one anticipates that spectrally resolving the
third-harmonic signal, generated by the sample, versus the time delay of two exciting pulses, gives additional insight.
Indeed, our experiments reveal that different spectral components of the third-harmonic signal can exhibit
substantially different temporal dynamics -- information that would obviously not be available from a spectrally
integrated experiment. The comparison of our experimental data with theory allows us to determine the dominant source
of the underlying optical nonlinearity. Finally, we conclude in Sec.\,\ref{SecConclus}.

\section{Nonlinear optics of ensembles of Lorentzian oscillators}\label{SecEnsembles}

To probe metal nanoparticles with diameters in the 10--200 nm range by linear- or nonlinear-optical techniques, one
often averages over several thousands of these particles in order to obtain an acceptable signal strength. Depending on
the fabrication method (e.g., lithographic patterning,\cite{CraigheadAPL84} Volmer-Weber growth \cite{VolmerZPC26}), a
distribution in particle size and shape results, leading to a distribution of plasmon-resonance
frequencies.\cite{KreibigBook} Ensemble linear-optical experiments alone cannot distinguish this inhomogeneous
contribution from the homogeneous linewidth (resulting from an expected plasmon decay time $\tau$ of a few
femtoseconds). Therefore, e.g., a combination of linear and nonlinear methods has to be used to extract both
homogeneous and inhomogeneous contributions. Nevertheless, not all nonlinear methods allow for this determination.

We begin by discussing analytic results for the limit of $\delta$ pulses and Lorentzian inhomogeneous broadening, and
continue with numerical simulations for finite Gaussian inhomogeneous broadening and pulses of finite duration.

\subsection{Analytic calculations}

Following along the lines of Ref.\,\onlinecite{LamprechtAPB99}, we start by describing the particle plasmon by an
oscillating particle with charge $q$, mass $m$, and displacement $x(t)$, driven by an electric field $E(t)$ via
\begin{equation}\label{EqSglOscDiffEq}
\ddot{x}+2\gamma\dot{x}+\Omega_0^2x +(\xi x^2+\eta x^3+...) = \frac{q}{m} E(t)\,.
\end{equation}
For the interferometric experiments to be described, $E(t)$ corresponds to a pair of copropagating pulses with time
delay $T$. In linear optics, i.e., for $\xi=\eta=...=0$, this leads to a Lorentz oscillator resonance at the damped
eigenfrequency $\Omega=\sqrt{\Omega_0^{2}-\gamma^2}$ with a half width at half maximum (HWHM) $\gamma=1/T_2=1/(2\tau)$,
the homogeneous linewidth. $T_2$ is the dephasing time. To first order in the laser electric field, the polarization
$P^{(1)}\propto x^{(1)}(t)$ is given by
\begin{equation}\label{EqSglOscLinPol}
x^{(1)}(t) \propto \Omega^{-1}\, \int_{-\infty}^{t} dt'\ \textrm{e}^{-\gamma (t-t')}\ \sin [\Omega ( t-t' )] \,
E(t')\,.
\end{equation}
Upon excitation with resonant pulses, the Fourier transform of $P^{(1)}(t)$ contains frequency components around
$\pm\Omega$. To second order in the laser electric field, $-\xi (x^{(1)}(t))^2$ is the driving term for the
second-order displacement $x^{(2)}(t)$. Provided that this driving term is off-resonant with respect to $\Omega$, we
obtain the second-order polarization $P^{(2)}(t) \propto x^{(2)}(t)$ with
\begin{equation}\label{EqSglOscNonlinPol}
x^{(2)}(t)\propto[ x^{(1)}(t) ] ^2\,.
\end{equation}
The Fourier transform of this expression contains frequency components around $\pm 2\Omega$, i.e., optical
second-harmonic generation, and components around zero frequency, i.e., optical rectification. Furthermore,
Ref.\,\onlinecite{LamprechtAPB99} argued (see also formulae in Refs.\,\onlinecite{LamprechtAPB97,TraegerAPB99}) that
the signal $S$ measured by a slow detector is given by the integral of the nonlinear intensity over time, i.e.,
\begin{equation}\label{EqIACIntTime}
S^{(2)}_{\rm SHG\,+\,OR}(T)\propto \int_{-\infty}^{\infty}dt\ [ P^{(2)}(t) ] ^2\,.
\end{equation}
It is crucial to note that this expression comprises both SHG and OR. This, however, is in contrast to what is actually
measured in a second-order interferometric autocorrelation (IAC) setup, where one selectively detects the SHG by means
of a photomultiplier tube behind {\it optical filters}, which suppress contributions other than SHG.\cite{remark}

For reasons of simplicity and to allow for analytic results, we first discuss excitation with a pair of $\delta$
pulses, i.e.,
\begin{equation}
E(t)=\tilde{E}_0\,[\delta(t)+\delta(t-T)]\,.
\end{equation}
It is clear from the symmetry that the nonlinear signals only depend on $|T|$. Thus, we only consider $T \geq 0$ in
what follows. For a single homogeneously broadened oscillator we obtain
\begin{eqnarray}\label{EqDeltaPulseLinPol}
P^{(1)}(t) & \propto & \Omega^{-1} \{ \Theta(t)\,\textrm{e}^{-\gamma t} \sin(\Omega t)
\nonumber\\
&& + \Theta(t-T)\,\textrm{e}^{-\gamma (t-T)} \sin[\Omega(t-T)]\}\,.
\end{eqnarray}
This leads to the second-order polarization
\begin{eqnarray}\label{EqDeltaPulseNonlinPol}
&&P^{(2)}(t)\nonumber\\
& \propto & \Omega^{-2} ( \Theta(t)\,\textrm{e}^{-2\gamma t}
[1-\cos(2\Omega t)] \nonumber\\
&& + \Theta(t-T)\,\textrm{e}^{-2\gamma (t-T)}
\{1-\cos[2\Omega(t-T)]\} \nonumber\\
&& + 2\Theta(t-T)\,\textrm{e}^{-\gamma (2t-T)} \{\cos(\Omega T )-\cos[\Omega(2t-T)]\})\,.\nonumber\\
&&
\end{eqnarray}
Let us now consider an inhomogeneously broadened ensemble of oscillators with fixed damping $\gamma$ and a Lorentzian
distribution of eigenfrequencies $\Omega$ with distribution function
\begin{equation}\label{EqLorWeigh}
\rho(\Omega)= \frac{\Gamma/\pi}{(\Omega-\overline{\Omega})^2+\Gamma^2}\,,
\end{equation}
which is centered around frequency $\overline{\Omega}$. The HWHM of this inhomogeneous distribution is $\Gamma$. To
work out the convolution, we approximate the prefactor $1/\Omega^{2}$ in Eq.\,(\ref{EqDeltaPulseNonlinPol}) by
$1/\overline{\Omega}^{2}$. This approximation is justified in the limit $\Gamma\ll\overline{\Omega}$, which is usually
well satisfied for lithographically fabricated particles. These two steps together lead to
\begin{eqnarray}\label{EqDeltaPulseConvNonlin}
P^{(2)}_{\rm inhom}(t) & \propto & \int_{-\infty}^{\infty} d\Omega\, \rho(\Omega)P^{(2)}(t) \nonumber\\[5pt]
&\propto & +\Theta(t)\,\textrm{e}^{-2\gamma t} \nonumber\\
&&+\Theta(t-T)\,\textrm{e}^{-2\gamma (t-T)} \nonumber\\
&&+2\Theta(t-T)\,\textrm{e}^{-\gamma (2t-T)-\Gamma T}
\cos(\overline{\Omega}T) \nonumber\\[5pt]
&&-\Theta(t)\,\textrm{e}^{-2(\gamma+\Gamma)t}
\cos(2\overline{\Omega}t) \nonumber\\
&&-\Theta(t-T)\,\textrm{e}^{-2(\gamma+\Gamma)(t-T)}
\cos[2\overline{\Omega}(t-T)] \nonumber\\
&&-2\Theta(t-T)\,\textrm{e}^{-(\gamma+\Gamma)(2t-T)} \cos[\overline{\Omega}(2t-T)]\,.\nonumber\\
&&
\end{eqnarray}
The first three lines correspond to OR, the last three lines to SHG. Note that the latter solely depends on the total
width $\gamma+\Gamma$. In linear optics, the width of the inhomogeneous ensemble results from the convolution of a
Lorentzian with homogeneous width $\gamma$ with a Lorentzian of inhomogeneous width $\Gamma$. This leads to a total
width of the resonance in linear optics of $\gamma+\Gamma$. Thus, both the linear response and the correctly calculated
SHG depend in the very same manner on the homogeneous and inhomogeneous linewidth, and a distinction is strictly not
possible.

In contrast, the contribution from OR does not simply depend on $\gamma+\Gamma$, potentially allowing for a distinction
between homogeneous and inhomogeneous linewidths. By {\it erroneously} including OR in the calculated interferometric
``SHG signal,'' one can {\it seemingly} separate the homogeneous and inhomogeneous contributions to the linewidth.

We have performed an analogous calculation for the third-order nonlinear-optical response. For third-harmonic
generation and $T\geq 0$, we find that the ensemble THG polarization is
\begin{eqnarray}\label{EqDeltaPulseConvTHG}
&&P^{(3)}_{\rm THG,\,inhom}(t)\nonumber\\
& \propto & -\Theta(t)\,\textrm{e}^{-3(\gamma+\Gamma)t}
\cos(3\overline{\Omega}t) \nonumber\\
&&-\Theta(t-T)\,\textrm{e}^{-3(\gamma+\Gamma)(t-T)}
\cos[3\overline{\Omega}(t-T)] \nonumber\\
&&-3\Theta(t-T)\,\textrm{e}^{-(\gamma+\Gamma)(3t-T)} \cos[\overline{\Omega}(3t-T)] \nonumber\\
&&-3\Theta(t-T)\,\textrm{e}^{-(\gamma+\Gamma)(3t-2T)} \cos[\overline{\Omega}(3t-2T)]\,.
\end{eqnarray}
The THG again only depends on $\gamma+\Gamma$, and no information on the homogeneous linewidth $\gamma$ can be
obtained. However, self-phase modulation would provide such information. In a non-copropagating geometry, the latter
would give rise to a diffracted four-wave-mixing signal. Corresponding calculations have been presented in
Ref.\,\onlinecite{WegenerPRA90}.

Broadly speaking, nonlinear-optical signals of the type $\omega+\omega$ (SHG) or $\omega+\omega+\omega$ (THG), etc. do
{\it not} allow one to distinguish between homogeneous and inhomogeneous contributions to the linewidth, whereas
signals of the type $\omega-\omega$ (OR) or $\omega+\omega-\omega$ (SPM or FWM), etc. do allow for such distinction.
The ``-'' sign in OR, SPM, FWM, etc., effectively reverses the time axis in analogy to phase conjugation. For example
in FWM, the ``-'' sign leads to the well-known photon-echo response.\cite{WegenerPRA90} At this point, a decay of the
ensemble polarization due to inhomogeneous broadening (just interference) can be reversed, whereas damping due to
homogeneous broadening (a dissipative process) cannot be reversed.

\subsection{Numerical calculations}

The presented analytical calculations for $\delta$ pulses are appropriate if the (complex) laser electric field
spectrum exhibits negligible variation on the scale of the homogeneous linewidth $\gamma$. For longer pulses, we have
performed numerical simulations. As described above, the correct way to calculate the SHG contribution is to spectrally
filter the second-order response of the oscillator ensemble. To obtain the final IAC signal as a function of the time
delay $T$, the square modulus of this filtered second-order polarization has to be integrated with respect to
frequency. To allow for a direct comparison with the results of Ref.\,\onlinecite{LamprechtAPB99}, we also use
sech$^2$-shaped 15\,fs pulses with a center wavelength of 780\,nm, resonantly exciting the ensemble. The latter has a
Gaussian distribution of resonances and is discretized in steps of 1\,nm.

\begin{figure}[tbp]
\centerline{\includegraphics[width=8.6cm,trim=0cm 0cm 0.7cm 0cm,clip]{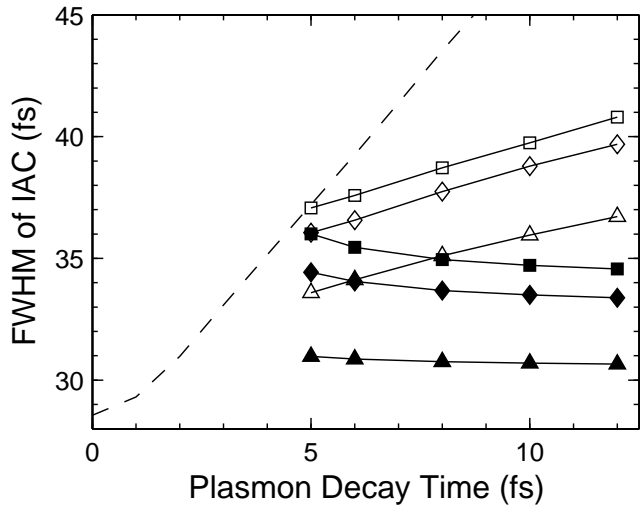}} \caption{\small
Recalculated data of Fig.\,2 of Ref.\,\onlinecite{LamprechtAPB99}. The dashed line describes the simulated FWHM of the
second-order interferometric autocorrelation (IAC) from a \emph{single} resonant Lorentz oscillator (the particle
plasmon) versus plasmon decay time $\tau=T_2/2$. The solid lines represent \emph{ensembles} of oscillators with
eigenfrequencies following a Gaussian distribution, for which the width is determined by fixing the total extinction
linewidth $\Delta\lambda$ (squares, $\Delta\lambda$=70\,nm; diamonds, $\Delta\lambda$=80\,nm; triangles,
$\Delta\lambda$=125\,nm). The autocorrelation width is indicated by the full symbols for considering \emph{only} the
contribution of SHG. Corresponding results for (erroneously) including SHG \emph{and} OR are shown by the open
symbols.} \label{FigLampCorr}
\end{figure}

Figure\,\ref{FigLampCorr} shows the resulting full width at half maximum (FWHM) of the interferometric autocorrelation
as a function of the plasmon decay time $\tau$. The full symbols correspond to the correct calculation, whereas the
open symbols erroneously comprise the OR contribution and qualitatively reproduce the results of
Ref.\,\onlinecite{LamprechtAPB99} (see their Fig.\,2). For each of the curves in our Fig.\,\ref{FigLampCorr}, the total
linewidth of the linear-optical spectrum is fixed. The squares, diamonds, and triangles correspond to a fixed
extinction linewidth (FWHM) of $\Delta \lambda=70$, $80$, and $125\,\rm nm$, respectively (see parameters of Fig.\,2 of
Ref.\,\onlinecite{LamprechtAPB99}). The dashed curve corresponds to a single (homogeneously broadened) oscillator for
reference. The correct results and those including the OR contribution differ strongly -- as in our analytical
calculations. In particular, the slopes of the correct curves in our Fig.\,\ref{FigLampCorr} are very nearly zero
(within typical experimental error bars of 1\,fs), while the incorrect simulations have a small positive slope. Thus,
using the correct curves one {\it cannot} infer the plasmon decay time from measured interferometric autocorrelations,
whereas the incorrect curves erroneously suggest this possibility.\cite{LamprechtAPB99} We conclude that, under
inhomogeneous conditions, the homogeneous linewidth cannot be determined by analyzing linewidths from linear optics and
SHG (or THG) measurements. Consequently, we will refrain from making any quantitative statements about plasmon decay
times from now on.

We note in passing that the IAC acquires artificial ``wings'' \cite{LamprechtAPB99} if the OR contribution is
erroneously included. Indeed, such wings are visible in Fig.\,1a of Ref.\,\onlinecite{LamprechtAPB99}. They disappear
in the correct calculation (not shown).

\section{Linear optics of two coupled Lorentzian oscillators}\label{SecLinOpt}

In this section, we discuss the linear-optical properties of two coupled Lorentz oscillators. As already mentioned in
the introduction, this system can serve as a simple model for metallic photonic crystal slabs. The results of this
Sec.\,can be compared with the linear-optical experiments (Sec.\,\ref{SecExpmts}) and, furthermore, are the basis for
our discussion of the nonlinear-optical properties in Sec.\,\ref{SecNonlinOpt}.

Generalizing Eq.\,(\ref{EqSglOscDiffEq}) to two coupled, oscillating particles of equal mass $m$ leads to
%\begin{widetext}
\begin{subequations}\label{Eq2OscDiffEqs}
\begin{eqnarray}
\ddot{x}_{\rm pl}\!+\!2\gamma_{\rm pl}\dot{x}_{\rm pl}\!+\!\Omega_{\rm pl}^2 x_{\rm pl}\!+\!({\rm NL})_{\rm
pl}\!-\!\Omega_{\rm c}^2
x_{\rm wg} & = & \frac{q_{\rm pl}}{m} E(t)\,, \nonumber\\
&&\\
\ddot{x}_{\rm wg}\!+\!2\gamma_{\rm wg}\dot{x}_{\rm wg}\!+\!\Omega_{\rm wg}^2 x_{\rm wg}\!+\!({\rm NL})_{\rm
wg}\!-\!\Omega_{\rm c}^2
x_{\rm pl} & = & \frac{q_{\rm wg}}{m} E(t)\,.\nonumber\\
&&
\end{eqnarray}
\end{subequations}
%\end{widetext}
Here, $x_{\rm pl}(t)$ and $x_{\rm wg}(t)$ are the displacements representing the plasmon and waveguide oscillations,
respectively. The resonance frequencies, (homogeneous) half widths at half maximum, and oscillator strengths of the
uncoupled system are denoted by $\Omega_{j}$, $\gamma_{j}$, and $q_{j}$ (${j}={\rm pl,wg}$), respectively. $\Omega_{\rm
c}^2$ represents the coupling strength between the oscillators. The nonlinear terms (denoted by $\rm NL$) are discussed
in Sec.\,\ref{SecNonlinOpt} and ignored here.

In order to make the resulting formulas transparent, we immediately discuss a few parameters in terms of their
experimentally relevant values. Since the uncoupled waveguide resonance is extremely sharp \cite{SharonJOSAA97} as
compared to the plasmon width, we set the waveguide damping $\gamma_{\rm wg}=0$. In the following, we derive formulas
for an arbitrary waveguide oscillator strength $q_{\rm wg}$; however, most aspects can already be understood in the
simpler case $q_{\rm wg}=0$. For typical sample parameters, $|q_{\rm wg}|\ll|q_{\rm pl}|$, i.e., the area under the
extinction curve of the (uncoupled) waveguide mode is much smaller than that of the plasmon.

In the frequency domain, Eqs.\,(\ref{Eq2OscDiffEqs}) can easily be solved analytically. For monochromatic excitation,
i.e., for $E(t)=\tilde{E}_0\,e^{-i\omega t}+{\rm c.c.}$, this leads to the first-order displacements
$x^{(1)}_{j}(t)=\tilde{x}^{(1)}_{j}(\omega)\, e^{-i\omega t}+{\rm c.c.}$ and the polarizations
$\tilde{P}^{(1)}_j(\omega) =N\,q_j\,\tilde{x}^{(1)}_j(\omega)$. $N$ is the density of the oscillators. Note that in the
case $q_{\rm wg}=0$, only $\tilde{x}^{(1)}_{\rm pl}(\omega)$ contributes to the polarization. The total linear
polarization becomes
\begin{widetext}
\begin{equation}\label{Eq2OscLinPol}
\tilde{P}^{(1)}(\omega)=\frac{N}{m}\, \frac{q_{\rm pl}^2(-\omega^2+\Omega_{\rm wg}^2) + 2q_{\rm pl}q_{\rm
wg}\Omega_{\rm c}^2 + q_{\rm wg}^2(-\omega^2-2i\omega\gamma_{\rm pl}+\Omega_{\rm pl}^2)}
{(-\omega^2-2i\omega\gamma_{\rm pl}+\Omega_{\rm pl}^2)(-\omega^2+\Omega_{\rm wg}^2)-\Omega_{\rm c}^4}\, \tilde{E}_0\,.
\end{equation}
The linear susceptibility $\tilde{\chi}^{(1)}(\omega)=\tilde{P}^{(1)}(\omega) / (\varepsilon_0 \tilde{E}_0)$ and the
absorption coefficient
\begin{equation}\label{Eq2OscAbsorption}
\alpha(\omega) = (\omega/c_0)\,\mathrm{Im}[\tilde{\chi}^{(1)}(\omega)] = \alpha_{\rm pl}\, \frac{4\gamma_{\rm
pl}^2\omega^2[\omega^2-\Omega_{\rm wg}^2- (q_{\rm wg}/q_{\rm pl})\Omega_{\rm c}^2 ]^2} {[(\omega^2-\Omega_{\rm pl}^2)
(\omega^2-\Omega_{\rm wg}^2)-\Omega_{\rm c}^4]^2\,+\,4\gamma_{\rm pl}^2\omega^2(\omega^2-\Omega_{\rm wg}^2)^2}
\end{equation}
\end{widetext}
immediately follow. $\varepsilon_0$ is the vacuum permittivity, $c_0$ the vacuum speed of light, and $\alpha_{\rm
pl}=Nq_{\rm pl}^2/(2m\varepsilon_0c_0\gamma_{\rm pl})$ the maximum absorption coefficient of the uncoupled plasmon
oscillation.

\begin{figure}[tbp]
\centerline{\includegraphics[width=8.6cm,trim=0cm 0cm 0.7cm 0cm,clip]{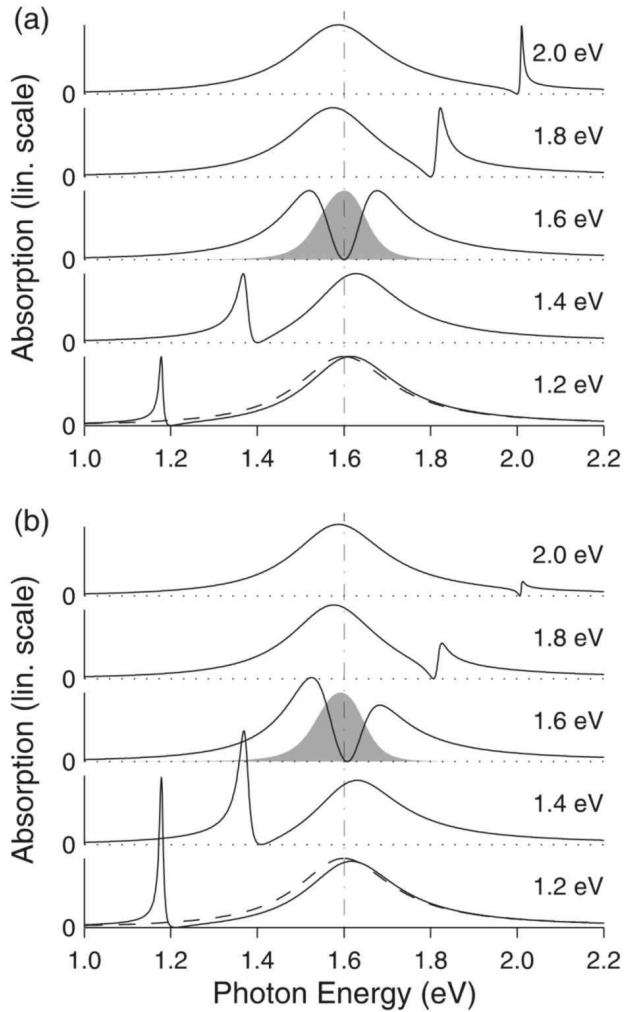}} \caption{\small Optical
absorption spectra according to Eq.\,(\ref{Eq2OscAbsorption}) (solid lines). One observes an anticrossing behavior when
varying the waveguide resonance frequency $\Omega_{\rm wg}$ with respect to the fixed plasmon resonance frequency
$\Omega_{\rm pl}$. Note the highly asymmetric, Fano-like lineshape of the peaks. All curves are displayed on the same
scale. $q_{\rm wg}/q_{\rm pl}$ equals 0 in (a) and $+0.1$ in (b). Common parameters are $\hbar\Omega_{\rm pl}=1.6\,\rm
eV$, $\hbar\Omega_{\rm wg}$ as denoted for each curve, $\gamma_{\rm pl}=1/(2\tau)$, $\tau=2.5\,\rm fs$, and
$\hbar\Omega_{\rm c}=0.5\,\rm eV$. The dashed lines represent the pure plasmonic absorption in the absence of coupling,
i.e., for $\hbar\Omega_{\rm c}=0$. The gray areas shown for $\Omega_{\rm wg}=1.6\,\rm eV$ depict the square modulus of
the waveguide amplitude, $|\tilde{x}^{(1)}_{\rm wg}|^2$, each exhibiting a single peak. In (a), the square modulus of
the plasmon amplitude, $|\tilde{x}^{(1)}_{\rm pl}|^2$, is roughly proportional to the corresponding absorption
spectrum. The vertical line is a guide to the eye. } \label{FigLinAbsTheo}
\end{figure}

Examples of absorption spectra are shown in Fig.\,\ref{FigLinAbsTheo}(a) for $q_{\rm wg}/q_{\rm pl}=0$ and
Fig.\,\ref{FigLinAbsTheo}(b) for $q_{\rm wg}/q_{\rm pl}=+0.1$. One obtains the anticipated anticrossing behavior. For
$q_{\rm wg}=0$, absorption maxima appear at the spectral positions
\begin{equation}\label{Eq2OscMaxFreqs}
\Omega_{\rm a,b}^2= \left(\Omega_{\rm pl}^2+\Omega_{\rm wg}^2\right)/2\ \pm\
        \left[\left(\Omega_{\rm pl}^2-\Omega_{\rm wg}^2\right)^2/4+\Omega_{\rm c}^4\right]^{1/2}\,.
\end{equation}
These positions coincide with the normal mode frequencies of the coupled, but undamped system. For small $\Omega_{\rm
c}$ and for $\Omega_{\rm pl}=\Omega_{\rm wg}$, the corresponding Rabi splitting is given by $\Omega_{\rm
c}^2/\Omega_{\rm pl}$. Hence, the two oscillators can be considered as ``resonant'' if $|\Omega_{\rm pl}-\Omega_{\rm
wg}|\ll \Omega_{\rm c}^2/\widehat{\Omega}$ with $\widehat{\Omega}=(\Omega_{\rm pl}+\Omega_{\rm wg})/2$, and as
``nonresonant'' otherwise. In contrast to frequent belief, the lineshapes in Fig.\,\ref{FigLinAbsTheo} do {\it not}
correspond to the sum of two {\it effective} Lorentz oscillators. One rather gets a highly asymmetric, Fano-like
lineshape. Usually, a Fano resonance results from the coherent interaction of a discrete quantum mechanical state with
a continuum of states.\cite{FanoPR61,Marburg} In our purely classical model, a single sharp oscillator coherently
interacts with a strongly broadened second oscillator. The latter replaces the continuum. One result of the Fano-like
interaction is that one obtains zero absorption between the two absorption maxima. The position of this zero appears at
the root of the numerator of (\ref{Eq2OscAbsorption}), i.e., at or near the spectral position of the (uncoupled)
waveguide mode $\Omega_{\rm wg}$. Intuitively, this minimum is a result of destructive interference, which effectively
suppresses the response of the two absorption ``channels,'' of which the polarizations have a phase difference near
$\pi$. This phase difference will also be important in nonlinear optics (see Sec.\,\ref{SecNonlinOpt}). When $q_{\rm
wg}$ is changed from zero to a nonzero value, the positions of the absorption extrema shift slightly, and the two peaks
exhibit different heights as an additional characteristic. A reduced absorption of the more waveguide-like channel
results, e.g., in the case $q_{\rm wg}/q_{\rm pl}>0$ and $\Omega_{\rm pl}<\Omega_{\rm wg}$ [see top curves in
Fig.\,\ref{FigLinAbsTheo}(b)].

We note that, e.g., for $q_{\rm wg}=0$, the total absorption (\ref{Eq2OscAbsorption}) can be rewritten as a sum of two
``Lorentzians,'' but with {\it strongly frequency-dependent} dampings. In the time domain, these frequency-dependent
dampings correspond to a non-Markovian (and non-exponential) decay. For $\Omega_{\rm a}<\Omega_{\rm wg}<\Omega_{\rm
b}$, one solution can be described by oscillator {\it a} with constant resonance frequency $\Omega_{\rm a}$ and
frequency-dependent damping
%\begin{widetext}
\begin{eqnarray}\label{Eq2OscReplacemtDampg}
\gamma_{\rm a}(\omega)&&\nonumber\\
=&\left\{
  \begin{array}{r@{,\ \ }l}
  \frac{\gamma_{\rm pl}\alpha_{\rm pl}}{2\alpha(\omega)}\left(1+\left[1-\frac{\alpha^2(\omega)\left(\omega^2-\Omega_{\rm a}^2\right)^2}
  {\alpha_{\rm pl}^2\gamma_{\rm pl}^2\omega^2}\right]^{1/2}\right) & \omega<\Omega_{\rm wg} \\
  \infty & \omega\geq\Omega_{\rm wg}\, \\
  \end{array}
\right.&\nonumber\\
&&
\end{eqnarray}
%\end{widetext}
and an analogous expression for the oscillator {\it b}.

\section{Nonlinear optics of two coupled Lorentzian oscillators}\label{SecNonlinOpt}

In this section, we discuss the nonlinear-optical properties of two coupled Lorentz oscillators in terms of
third-harmonic generation. We consider an inversion-symmetric medium, hence all second-order nonlinear terms in
Eqs.\,(\ref{Eq2OscDiffEqs}) are zero. At first sight, one might only expect third-order nonlinear terms like $({\rm
NL})_{\rm pl} \propto x_{\rm pl}^3$ or $({\rm NL})_{\rm wg} \propto x_{\rm wg}^3$ in Eqs.\,(\ref{Eq2OscDiffEqs}).
Mathematically, the most general form is given by the terms
\begin{equation}\label{EqNonLinTermsDiffEqs}
\eta_{j,k}[x_{\rm pl}(t)]^{3-k}[x_{\rm wg}(t)]^{k}
\end{equation}
appearing in the differential equation for $x_j(t)$, respectively (${j}={\rm pl,wg}; k=0,1,2,3$). Here, we are only
interested in THG, which is off-resonant. In a perturbational approach the THG contributions to the third-order
displacements are given by
\begin{equation}\label{EqNonLinTermsDispl}
x^{(3)}_j(t) \propto \sum_{k=0}^3 \eta_{j,k}[x^{(1)}_{\rm pl}(t)]^{3-k}[x^{(1)}_{\rm wg}(t)]^{k}\,.
\end{equation}
The eight parameters $\eta_{j,k}$ can be reduced to four, i.e., $\eta_k=\sum_j q_j \eta_{j,k}$ with $k=0,1,2,3$, because the optical polarization
is given by the weighted sum of the displacements.
This immediately leads to the following general form for the THG contribution to the third-order polarization
\begin{equation}\label{EqNonLinTermsPol4}
P^{(3)}(t) \propto \sum_{k=0}^3 \eta_{k}[x^{(1)}_{\rm pl}(t)]^{3-k}[x^{(1)}_{\rm wg}(t)]^{k}\,.
\end{equation}
We note in passing that this form is generally different from the ansatz $P^{(3)}(t) \propto  [P^{(1)}(t)]^3$ (in
analogy to Ref.\,\onlinecite{ZentgrafPRL04}), which leads to $\eta_k\propto{3\choose k}\,q_{\rm pl}^{3-k} q_{\rm
wg}^k$.

For the numerical computation of THG spectra, we start off in the time domain. $E(t)$ is chosen \cite{OurPulses} to
resemble the 5\,fs laser pulses of the experiments (5\,fs Gaussian pulses deliver qualitatively similar results for all
conditions discussed below). Furthermore, we fix $\Omega_{\rm pl}=1.67\,\rm eV$, $\Omega_{\rm wg}=1.56\,\rm eV$,
$\Omega_{\rm c}=0.66\,\rm eV$, $\tau=1.06\,\rm fs$, and $q_{\rm wg}/q_{\rm pl}=+0.085$. These parameters correspond to
sample {\it A} in Sec.\,\ref{SecExpmts}, which can be considered as ``resonant'' according to the definition given in
Sec.\,\ref{SecLinOpt}. Integration of Eqs.\,(\ref{Eq2OscDiffEqs}) yields the first-order displacements $x^{(1)}_j(t)$
and, with (\ref{EqNonLinTermsPol4}), the third-order polarization. The square modulus of its filtered Fourier transform
delivers the THG intensity spectrum. Spectra are calculated as a function of the spectrometer photon energy and the
time delay between the two excitation pulses, $T$.

\begin{figure}[hp]
\centerline{\includegraphics[width=8.5cm,trim=0cm 0cm 0.7cm 0cm,clip]{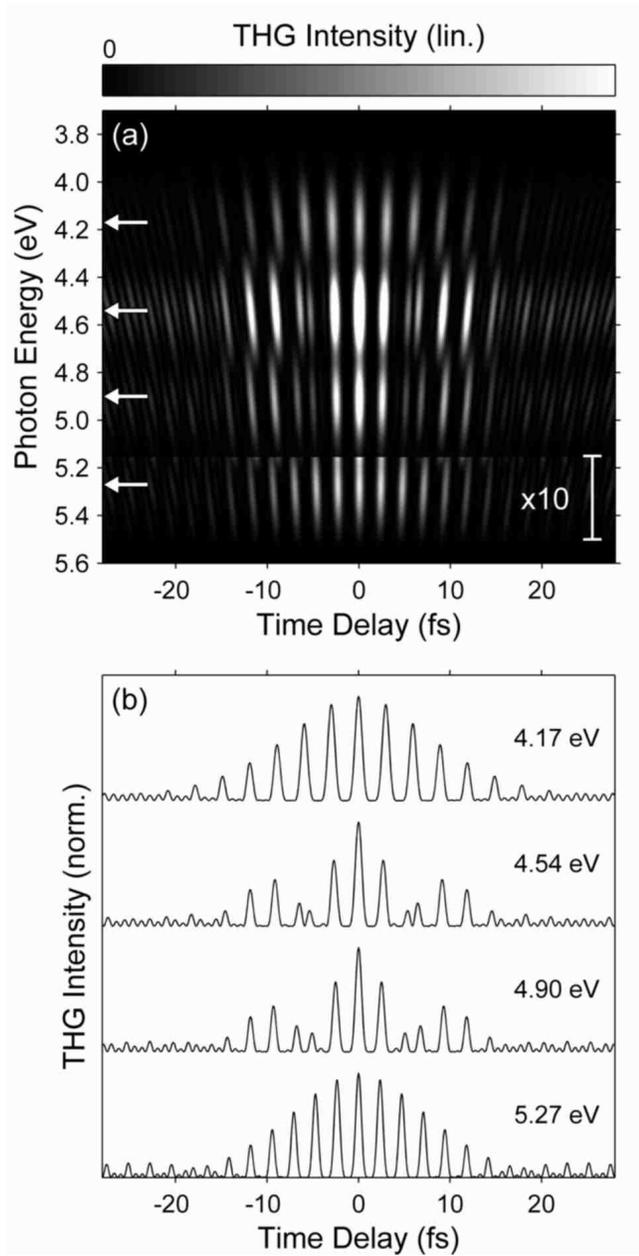}} \caption{\small (a)
Optical THG intensity derived from the coupled nonlinear oscillators. The THG intensity is shown on a saturated gray
scale, versus spectrometer photon energy and time delay $T$ between the two excitation pulses. At $T=0$, the THG
spectrum exhibits four distinct peaks (the high-energy peak is amplified by a factor of 10 for the sake of clarity).
The four peaks exhibit different temporal behaviors. Corresponding cuts at the spectral peak positions indicated by the
white arrows in (a) are shown in (b). For better comparison, the curves are normalized to the same maximum and are
vertically displaced. Obviously, the first and fourth curves both have a smoothly decaying (upper) envelope, while only
the second and third curves show an envelope resulting from a beating. The nonlinearity parameters used are $\eta_k
\propto \delta_{k,0}$. The other parameters are quoted in the text. Compare with the corresponding experiment
(Fig.\,\ref{FigTHGExpmtC4} below). } \label{FigTHGTheoC4Pl3}
\end{figure}

\begin{figure}[hp]
\centerline{\includegraphics[width=8.5cm,trim=0cm 0cm 0.7cm 0cm,clip]{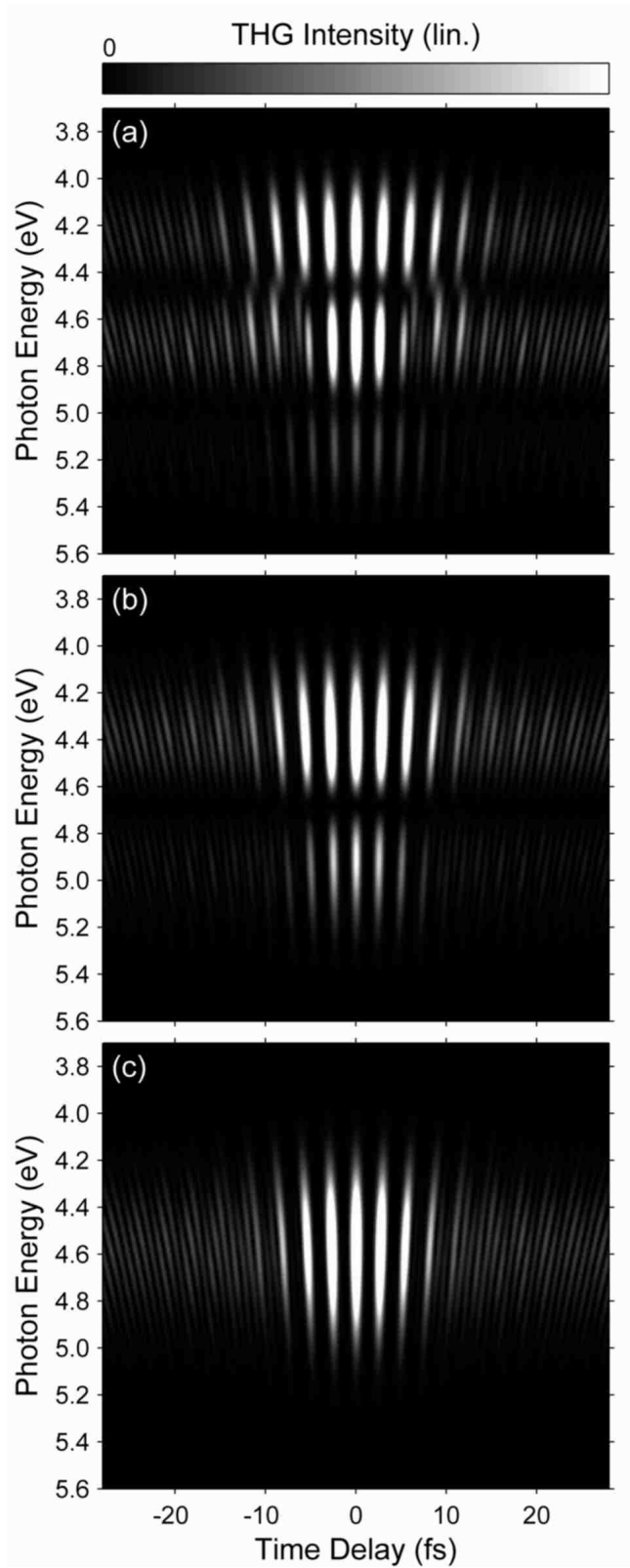}} \caption{\small As
Fig.\,\ref{FigTHGTheoC4Pl3}(a), but for different nonlinearity parameters. The shown THG contributions result from a
nonzero value of (a) $\eta_1$, (b) $\eta_2$, and (c) $\eta_3$ while keeping the other nonlinearity parameters zero. At
$T=0$, the number of spectral peaks is three, two, and one, respectively. (a), (b), and (c) are displayed on individual
gray scales. } \label{FigTHGTheoC4PlWg}
\end{figure}

We first discuss the case $\eta_k \propto \delta_{k,0}$ ($\delta_{k,l}$ is the Kronecker symbol). The corresponding
data set is shown in Fig.\,\ref{FigTHGTheoC4Pl3}(a). A cut at $T=0$ (not shown) reveals four broad but clearly distinct
spectral peaks in the THG spectrum. The appearance of four peaks can easily be understood in the frequency domain,
since the third-order polarization for this case is proportional to the twofold convolution of the displacement
$\tilde{x}^{(1)}_{\rm pl}(\omega)$ with itself, this displacement containing two peaks [see, e.g.,
Fig.\,\ref{FigLinAbsTheo}(a)]. The relative weights of the four peaks can be estimated employing the time domain.
Assuming $\delta$ pulses, $q_{\rm wg}=0$, and neglecting damping, the two {\it effective} oscillators (see previous
section) have comparable amplitude, and the general form of the THG polarization is proportional to
$\hphantom{\ref{FigTHGTheoC4PlWg}}$
\begin{eqnarray}\label{EqTHGRatio1331}
[\cos(\Omega_{\rm a}t)+\cos(\Omega_{\rm b}t)]^3 & \propto & ...\nonumber\\
&& +\cos(3\Omega_{\rm a}t)\nonumber\\
&& +3\cos[(2\Omega_{\rm a}+\Omega_{\rm b})t]\nonumber\\
&& +3\cos[(\Omega_{\rm a}+2\Omega_{\rm b})t]\nonumber\\
&& +\cos(3\Omega_{\rm b}t)\,.
\end{eqnarray}
This contains terms at three times the normal mode frequencies $\Omega_{\rm a}$ and $\Omega_{\rm b}$ as well as
\emph{spectral mixing products}. The relative amplitudes 1:3:3:1 of the frequency components $3\Omega_{\rm a}$,
$2\Omega_{\rm a}+\Omega_{\rm b}$, $\Omega_{\rm a}+2\Omega_{\rm b}$, and $3\Omega_{\rm b}$ lead to the intensity ratios
1:9:9:1. This means that the two central frequency components are more prominent, in agreement with the numerical
findings in Fig.\,\ref{FigTHGTheoC4Pl3}(a).

The pronounced dips between the four spectral peaks are closely related to the Fano-like lineshapes discussed in
Sec.\,\ref{SecLinOpt}. In linear optics, the phase relation between the two effective oscillators (absorption
``channels'') leads to destructive interference, and hence to zero absorption in the dip. The same destructive
interference is also responsible for the deep dips in the THG spectra.

The behavior of the THG intensity as a function of time delay $T$ differs among the four spectral peaks. Corresponding
cuts at the spectral peak positions indicated by the white arrows in Fig.\,\ref{FigTHGTheoC4Pl3}(a) are shown in
Fig.\,\ref{FigTHGTheoC4Pl3}(b). The curves exhibit the usual oscillations with the respective fundamental and harmonic
frequencies, enclosed in the (upper) envelope of interest. The first and fourth curves clearly show a smoothly decaying
envelope for increasing $|T|$. In contrast, the envelopes of the central two curves (which are associated with the
\emph{spectral mixing products}) reveal a beating. In spectrally integrated measurements,\cite{ZentgrafPRL04} this
distinction is not possible.

So far, we have only discussed the case $\eta_k \propto \delta_{k,0}$. Next, we calculate corresponding THG spectra for
different nonlinearity parameters (Fig.\,\ref{FigTHGTheoC4PlWg}). In each part of this figure, all nonlinear parameters
are zero except for a single one. The parts (a), (b), and (c) result from a nonzero value of $\eta_1$, $\eta_2$, and
$\eta_3$, showing three peaks, two peaks, and one peak, respectively. In the frequency domain this can again be
understood by the corresponding convolutions. Remember that $\tilde{x}^{(1)}_{\rm pl}(\omega)$ contains two peaks for
the values chosen here, whereas $\tilde{x}^{(1)}_{\rm wg}(\omega)$ only contains one peak (refer to gray areas in
Fig.\,\ref{FigLinAbsTheo}).

In general, all parameters $\eta_k$ can have nonzero values simultaneously. When adding up the nonlinear contributions
to the polarization, interference can result in a THG intensity with amplified or suppressed spectral peaks and dips,
spectrally shifted peak positions, or even with new peaks or dips which are not present at all in
Figs.\,\ref{FigTHGTheoC4Pl3}(a) and \ref{FigTHGTheoC4PlWg}. We will not go into a detailed analysis. We only note that
for $\eta_0\gg\eta_1>\eta_2=\eta_3=0$, the tendency is to suppress the high-energy peaks compared to the case
$\eta_0\neq 0$ and $\eta_1=\eta_2=\eta_3=0$.

The key feature of the calculations presented so far is that the THG spectra depend on the underlying source of the
nonlinearity, i.e., they depend on which of the coefficients $\eta_k$ is nonzero. In other words, observing four,
three, two, or just one peak in experimental THG spectra allows one to learn something about the system by comparison
with theory. This, however, is only possible for a certain regime of coupling between the two oscillators, which we
shall refer to as the regime of ``moderate coupling.'' Obviously, for very small coupling strengths, i.e., for small
values of $\Omega_{\rm c}$, the four spectral peaks in the THG spectrum of the case $\eta_k=\delta_{k,0}$ (discussed
above) merge into a single peak. In the other limit, i.e., for large values of $\Omega_{\rm c}$, also
$\tilde{x}^{(1)}_{\rm wg}(\omega)$ exhibits several peaks (unlike the gray areas in Fig.\,\ref{FigLinAbsTheo}), which
can, for example, lead to several spectral peaks in the THG spectrum for the case $\eta_k=\delta_{k,3}$ as well. By
numerical calculations for the ``resonant'' case (i.e., $\Omega_{\rm pl}=\Omega_{\rm wg}=\widehat{\Omega}$), for
$q_{\rm wg}=0$, and assuming $\delta$ pulses, we can specify the regime of ``moderate coupling'' by the condition
$0.15<\Omega_c^2/(2\gamma\widehat{\Omega})<1.35$. \emph{Thus, if one wants to learn something from the comparison of
experiment and theory, the coupling parameter of a sample has to be tailored correspondingly.}

\section{Experiments}\label{SecExpmts}

\begin{figure}[tbp]
\centerline{\includegraphics[width=8.6cm]{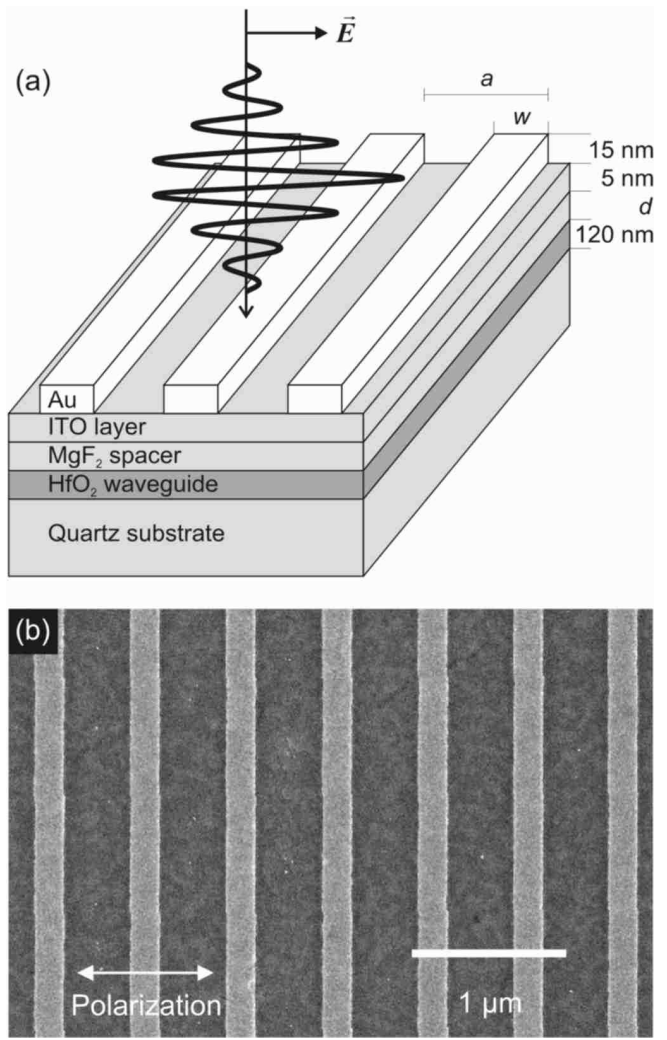}} \caption{\small (a) Scheme showing gold nanowires on top of
a set of dielectric layers forming the slab waveguide. HfO$_2$ is used as a high-index material, while MgF$_2$ serves
as a low-index spacer. The polarization of the normally incident white light or laser pulses is perpendicular to the
wires (TM polarization) for the experiments shown in Figs.\,\ref{FigExtcnExpmt}, \ref{FigTHGExpmtC4}, and
\ref{FigTHGExpmtF4}. Samples with different lattice constant $a$, wire width $w$, and spacer thickness $d$ are
investigated. (b) Scanning electron micrograph of the gold nanowires (light gray) on top of the waveguide (dark gray).
} \label{FigSampleGeom}
\end{figure}

The system of interest, a metallic photonic crystal slab, is schematically shown in Fig.\,\ref{FigSampleGeom}(a). The
coupling strength $\Omega_{\rm c}^2$ between the particle plasmon resonance and the Bragg resonance (waveguide mode)
can conveniently be tailored by the spacer thickness $d$. It is clear that an increasing spacer thickness leads to
decreasing coupling. We experimentally find that when choosing $d=30\,\rm nm$, the samples are within the regime of
``moderate coupling'' defined in the previous section, with a normalized coupling strength of
$\Omega_c^2/(2\gamma\widehat{\Omega})=0.43$ for sample {\it A} and $0.28$ for sample {\it B} (see below). These two
selected samples are presented in the following as examples for the ``resonant'' and ``nonresonant'' cases,
respectively (see definition in Sec.\,\ref{SecLinOpt}).

\subsection{Sample fabrication and linear-optical experiments}

First, the dielectric layers shown in Fig.\,\ref{FigSampleGeom}(a) are deposited in a high-vacuum chamber at pressures
around $10^{-6}\,\rm mbar$ via electron-beam evaporation. We use hafnium dioxide (HfO$_2$) as the high-index material
($n=1.95$) forming the core of the slab waveguide between the quartz substrate ($n=1.46$) and the magnesium fluoride
spacer layer (MgF$_2$, $n=1.38$). These dielectrics have been chosen for their transparency in the total spectral range
of interest as well as for minimum THG generation from the dielectric layers (as we investigated in independent
experiments). The 5-nm-thin indium tin oxide layer (ITO, $n=1.9$) is necessary to avoid charging effects in the
electron-beam writing process. Next, a photoresist layer is spun onto the sample, exposed by means of electron-beam
lithography, and developed. Finally, a 15-nm-thin gold film is evaporated and the metal on the remaining photoresist
areas is lifted-off. Each of the resulting gold nanowire arrays covers a total area of $(60\,\rm \mu m)^2$. The
electron micrograph in Fig.\,\ref{FigSampleGeom}(b) shows an enlarged view of a typical sample, revealing the high
quality of the resulting structures. Typically, we fabricate entire sets of arrays on one glass substrate. In such a
set, e.g., the lattice constant $a$ is varied from 500 to 650\,nm in steps of 25\,nm, and the nominal wire width from
around 120 to around 220\,nm in steps of 20\,nm. In this fashion, we fabricate and investigate a total of 42 nanowire
arrays on each substrate.

\begin{figure}[tbp]
\centerline{\includegraphics[width=8.6cm,trim=0cm 0cm 0.3cm 0cm,clip]{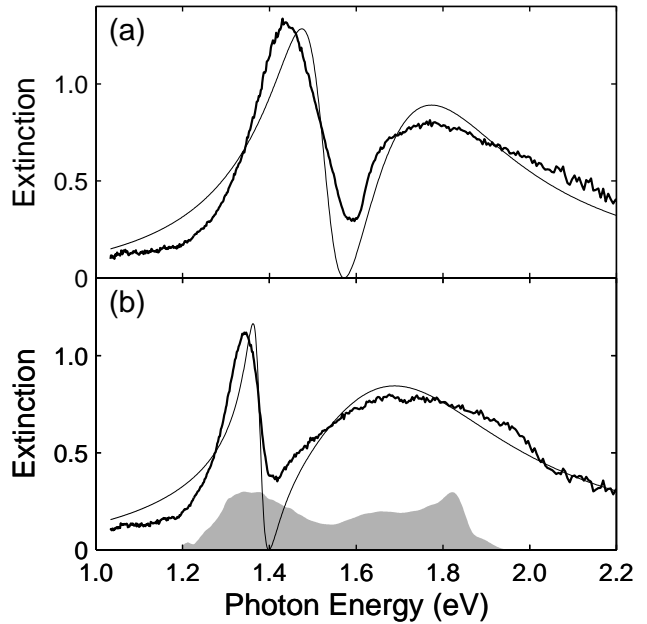}} \caption{\small Extinction
of the selected samples (a) {\it A} and (b) {\it B}. The thick curves show the extinction measured with a white-light
source, referenced to the substrate without gold structures. The thin lines are absorption spectra derived from the
model of coupled Lorentz oscillators, Eq.\,(\ref{Eq2OscAbsorption}), obtained by a nonlinear least-squares fit to the
corresponding experimental data. The gray area in (b) depicts the electric field spectrum (square root of the measured
intensity spectrum) of the laser pulses used for the nonlinear-optical experiments. } \label{FigExtcnExpmt}
\end{figure}

To connect to theory, the measured extinction spectra (negative logarithm of the intensity transmittance, referenced to
the substrate without gold structures) for TM polarization and for normal incidence are compared with
Eq.\,(\ref{Eq2OscAbsorption}) derived in Sec.\,\ref{SecLinOpt} (see Fig.\,\ref{FigExtcnExpmt}). In these experiments we
use a white-light source focused with a numerical aperture (NA) of about 0.025. Using a yet smaller NA tends to make
the extinction dip even more pronounced. A careful discussion of this aspect can be found in
Ref.\,\onlinecite{ChristPRB04}. We find a good qualitative agreement of our simple theoretical model and the
experiments. From a least-squares fit of the theory to the experiment (see Fig.\,\ref{FigExtcnExpmt}) we obtain all
relevant parameters, leaving only the nonlinear coefficients $\eta_k$ as free parameters for the nonlinear-optical
experiments to come. The experimental parameters of sample {\it A} ({\it B}) are: $a=550\,\rm nm$, $w=185\pm 5\,\rm nm$
($a=625\,\rm nm$, $w=195\pm 5\,\rm nm$). The fit parameters of sample {\it A} ({\it B}) are: $\hbar\Omega_{\rm
pl}=1.67\,\rm eV$, $\hbar\Omega_{\rm wg}=1.56\,\rm eV$, $\hbar\Omega_{\rm c}=0.66\,\rm eV$, $\tau=1.06\,\rm fs$, and
$q_{\rm wg}/q_{\rm pl}=+0.085$ ($\hbar\Omega_{\rm pl}=1.65\,\rm eV$, $\hbar\Omega_{\rm wg}=1.39\,\rm eV$,
$\hbar\Omega_{\rm c}=0.54\,\rm eV$, $\tau=0.97\,\rm fs$, and $q_{\rm wg}/q_{\rm pl} =+0.049$), where
$\tau=1/(2\gamma)$. The additional fit parameter $\alpha_{\rm pl}$, together with the coefficients $\eta_k$, determines
the absolute strength of the THG signals. Sample {\it A} is resonant, i.e., $\hbar|\Omega_{\rm pl}-\Omega_{\rm wg}| =
0.11\,{\rm eV} < 0.27\,{\rm eV} = \hbar\Omega_{\rm c}^2/ \widehat{\Omega}$, while sample {\it B} is nonresonant, i.e.,
$\hbar|\Omega_{\rm pl}-\Omega_{\rm wg}| = 0.26\,{\rm eV} > 0.19\,{\rm eV} = \hbar\Omega_{\rm c}^2/\widehat{\Omega}$
(see discussion in previous section).

\subsection{Nonlinear-optical experiments}

In our THG experiments, we use 5\,fs laser pulses derived from a laser system closely similar to the one described in
Ref.\,\onlinecite{MorgnerOL99} ($81\,\rm MHz$ repetition frequency). The pulses are sent into a Michelson
interferometer, which is actively stabilized by means of the ``Pancharatnam screw.''\cite{WehnerOL97} The linearly
polarized pulses emerging from the interferometer are focused onto the samples (normal incidence and TM polarization)
by a spherical mirror with a focal length of $f=100\,\rm mm$. To estimate the intensities in the spot and to determine
the effective NA, we have measured the spot size and the Rayleigh length by a knife-edge method in the horizontal
(vertical) direction. The determined spot radius of $12.5\,(12.3)\,\rm \mu m$ is significantly smaller than the size of
the nanowire arrays and leads to a pulse intensity around $I=3.9\times 10^{10}\,\rm W/cm^2$ and a laser fluence of
$200\,\rm \mu J/cm^2$ used in the experiments described below (for an average power of 80\,mW in front of the sample).
We will argue later that this fluence is still within the third-order perturbational limit. The corresponding Rayleigh
lengths of $490\,(560)\,\mu$m lead to an effective NA of 0.025 (0.022).

\begin{figure}[tbp]
\centerline{\includegraphics[width=8.6cm,trim=0cm 0cm 0.3cm 0cm,clip]{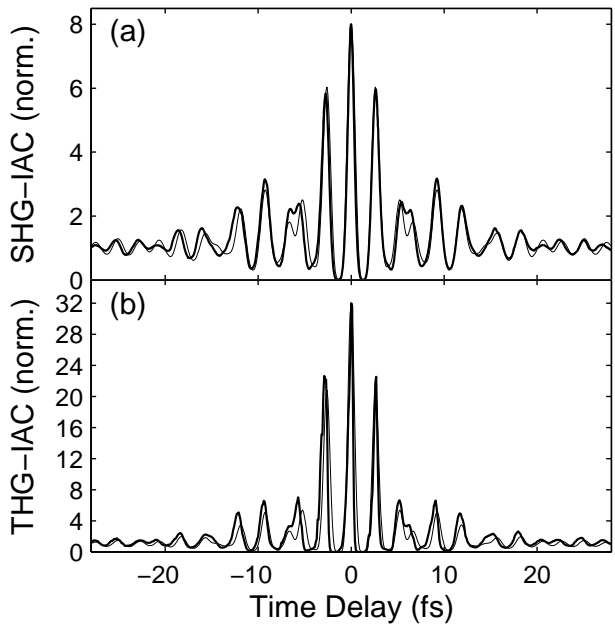}} \caption{\small
Characterization of the $5\,\rm fs$ laser pulses. (a) Second-order interferometric autocorrelation (SHG-IAC) measured
directly (thick line) and calculated (thin line) from the measured pulse spectrum under the assumption of a flat
spectral phase. (b) Third-order interferometric autocorrelation (THG-IAC) measured directly (thick line) and calculated
(thin line) in analogy to (a).} \label{FigIACPulses}
\end{figure}

In Fig.\,\ref{FigIACPulses} we depict the characterization of the laser pulses. The thick curve in (a) shows the usual
second-order autocorrelation obtained from a very thin $\beta$-barium-borate SHG crystal. The thin line is the
autocorrelation as calculated from the measured laser spectrum [see gray area in Fig.\,\ref{FigExtcnExpmt}(b)] under
the assumption of a spectrally flat phase. The good agreement indicates that the residual chirp on the 5\,fs pulses is
of minor importance. The thick curve in Fig.\,\ref{FigIACPulses}(b) shows the third-order autocorrelation function
measured via THG from the surface of a thick sapphire plate. The thin curve in (b) is the corresponding calculated
response under the same assumptions as in (a). Again, the agreement is very good. Notably, the envelope of the THG
signal has decayed by a factor of 4 for time delays of just two cycles of light. These curves in (b) can be considered
as the apparatus function and have to be compared with the measurements to be discussed in what follows. The achieved
ratio of 32:1 between the THG signal at zero time delay and large time delays, respectively, indicates good alignment
of the interferometer.

In the THG experiments, the emission from the samples in the forward direction is collected by another spherical mirror
(focal length $f=100\,\rm mm$), spectrally prefiltered by means of four fused-silica Brewster-angle prisms to suppress
the overwhelming fundamental laser light and spectrally resolved using a 0.5-m-focal-length grating spectrometer (with
a grating blazed at 250\,nm wavelength) connected to a uv-sensitive, back-illuminated, liquid-nitrogen-cooled
charge-coupled-device camera.

\begin{figure}[tbp]
\centerline{\includegraphics[width=8.5cm,trim=0cm 0cm 0.7cm 0cm,clip]{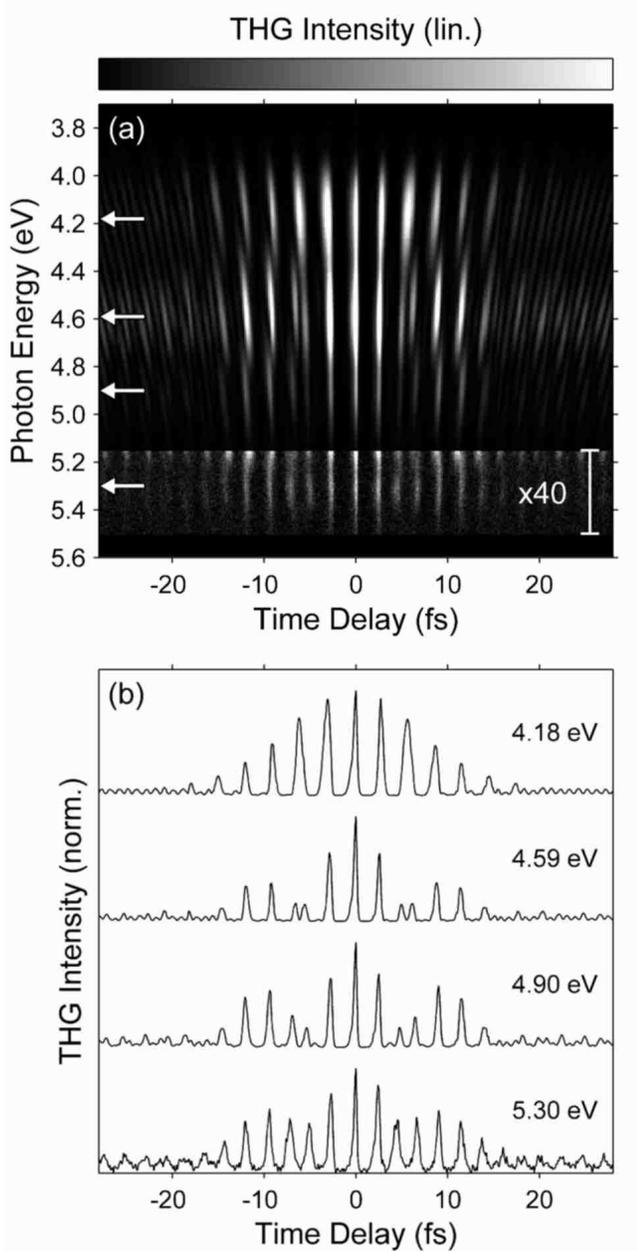}} \caption{\small (a)
Measured THG intensity from sample {\it A}. The THG intensity is shown on a saturated gray scale, versus spectrometer
photon energy and interferometric time delay. Near time delay $T=0$, the THG spectrum exhibits four peaks (the weak
high-energy peak is amplified by a factor of 40 to become visible). The four peaks exhibit a different temporal
behavior. Corresponding cuts at the spectral peak positions indicated by the white arrows in (a) are shown in (b). For
better comparison, the curves are normalized to the same maximum and are vertically displaced. The appearance and
absence of beating is discussed in the text. Compare with the corresponding theory (Fig.\,\ref{FigTHGTheoC4Pl3}). }
\label{FigTHGExpmtC4}
\end{figure}

Figure\,\ref{FigTHGExpmtC4} shows a typical data set of sample {\it A}, containing 600 individual spectra obtained in a
total of about 8\,min acquisition time. Here, the THG signal is plotted on a linear gray scale as a function of
spectrometer photon energy and time delay. The exact same representation has already been employed in the theory
Sec.\,(see Figs.\,\ref{FigTHGTheoC4Pl3} and \ref{FigTHGTheoC4PlWg}). Indeed, the linear-optical parameters of
Figs.\,\ref{FigTHGTheoC4Pl3} and \ref{FigTHGTheoC4PlWg} correspond to those of sample {\it A} [compare linear spectra
in Fig.\,\ref{FigExtcnExpmt}(a)]. Obviously, the measured nonlinear-optical spectra are much closer to those in
Fig.\,\ref{FigTHGTheoC4Pl3} than to any of Fig.\,\ref{FigTHGTheoC4PlWg}. In particular, four peaks occur in the spectra
at zero time delay. Also, the dependencies of the different spectral cuts versus time delay in
Fig.\,\ref{FigTHGExpmtC4}(b) closely resemble those in Fig.\,\ref{FigTHGTheoC4Pl3}(b). Again, the envelopes of the
first and fourth cuts show hardly any beating, whereas the envelopes of the second and third cuts reveal a pronounced
beating behavior [note that the very weak fourth peak in Fig.\,\ref{FigTHGExpmtC4}(a) spectrally overlaps with the wing
of the third peak, resulting in a small residual beating]. This comparison allows us to conclude that the nonlinear
model where only $\eta_0$ is nonzero is the appropriate one. {\it This means that the nonlinearity predominantly
originates from the particle plasmon} -- which is not {\it a priori} clear. This finding is consistent with the
experimental finding that the nonlinear signal decreases by a factor of 19 (and the multi-peak features disappear) when
going from the TM polarization used so far to the TE polarization. It is also consistent with the fact that the THG
signals drop by a factor of about 20 when going from the gold nanowire arrays to areas of the glass substrate where
only the dielectric layers are present.

It is important for our interpretation that the experiments are performed in the third-order perturbation regime --
which is also assumed in the theoretical analysis. Higher-order contributions would obviously modify the ratio of 32:1
between the THG signal at zero time delay and that at large time delays. In the experiments, the ratio of 32:1 is
reached within experimental uncertainty: From analyzing the upper (lower) envelope of spectrally integrated data like
those shown in Fig.\,\ref{FigTHGExpmtC4} but for time delays up to $\pm 60\,\rm fs$, we derive a ratio of 24:1 (35:1).
The actual ratio -- which refers to a comparison between zero and infinite time delay -- must lie between these two
ratios.

\begin{figure}[tbp]
\centerline{\includegraphics[width=8.5cm,trim=0cm 0cm 0.7cm 0cm,clip]{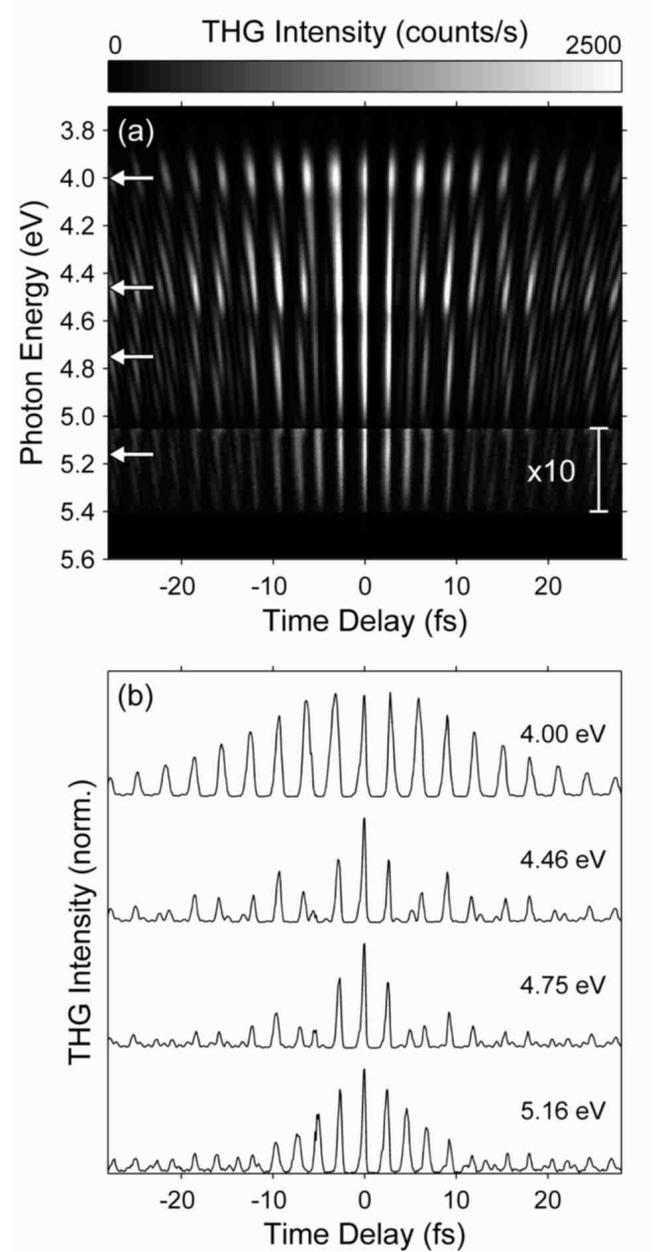}} \caption{\small (a)
Measured optical THG intensity from sample {\it B}. Near $T=0$, the THG spectrum clearly exhibits four peaks (the
high-energy peak is amplified by a factor of 10). Normalized cuts at the spectral peak positions indicated by the white
arrows in (a) are shown in (b). Note the much slower, smooth decay of the envelope of the cut at $4.00\,\rm eV$ as
compared to the cut at $5.16\,\rm eV$, and the beating, which only occurs in the other two curves. }
\label{FigTHGExpmtF4}
\end{figure}

In Fig.\,\ref{FigTHGExpmtF4}, we show the data set for the ``nonresonant'' sample {\it B}. As for sample {\it A}, four
spectral peaks are visible in the THG spectra. In contrast, however, the peaks in Fig.\,\ref{FigTHGExpmtF4}(a) have
rather different spectral widths, as expected from the fact that the two effective extinction peaks [see
Fig.\,\ref{FigExtcnExpmt}(b)] exhibit rather different spectral widths and our discussion of Sec.\,\ref{SecNonlinOpt}.
The different spectral widths in Fig.\,\ref{FigTHGExpmtF4}(a) correspond to strongly different decay times of the
envelopes in Fig.\,\ref{FigTHGExpmtF4}(b). Again, only the envelopes of the first and fourth cuts show a smooth decay,
whereas the envelopes of the second and third cuts exhibit a pronounced beating.

\section{Conclusions}\label{SecConclus}

We have investigated the linear- and nonlinear-optical lineshapes of metal nanoparticles and metallic photonic crystal
slabs.

For particle ensembles, we have shown analytically and numerically that the comparison of time-resolved femtosecond
second- or third-harmonic-generation experiments and extinction measurements does not allow one to distinguish between
homogeneous and inhomogeneous contributions to the linewidth. Optical-rectification or four-wave-mixing experiments
would provide such information.

For metallic photonic crystal slabs, we have demonstrated that the model of two coupled Lorentz oscillators describes
very well the key experimental features of linear optics. In particular, Fano-like lineshapes appear in the absorption
spectra. With regard to nonlinear optics, we have shown that -- within the regime of ``moderate coupling'' -- the
nonlinear-optical third-harmonic-generation spectra provide information on the underlying source of the optical
nonlinearity. Furthermore, the calculated nonlinear spectra reveal a beating in the spectral mixing products of the two
peaks from linear optics, but not in the third harmonics of the latter peaks.

Our corresponding experiments on third-harmonic generation of metallic photonic crystal slabs go beyond previous work
regarding improved temporal resolution and the fact that we spectrally resolve the interferometric third-harmonic
signal. The spectra reveal a distinct behavior of the various spectral components versus time delay. Some spectral
components exhibit a beating, others do not. Furthermore, the decay times of the envelopes strongly depend on the
spectral component. The measured spectra agree qualitatively very well with the predictions of the simple theoretical
model. The comparison allows us to identify the particle plasmon oscillation as the main source of nonlinearity.

\begin{acknowledgments}
We acknowledge support by the Center for Functional Nanostructures (CFN) of the Deutsche Forschungsgemeinschaft (DFG)
within subproject A1.5. The research of M.W. is further supported by Project No. DFG-We 1497/9-1. We thank  H.~Giessen,
J.~Kuhl, T.~Zentgraf, and A.~Christ for discussions.
\end{acknowledgments}

\end{document}